\documentstyle[aps]{revtex}
\begin{document}
\draft
\twocolumn[\hsize\textwidth\columnwidth\hsize\csname 
           @twocolumnfalse\endcsname
\title{Gravitational waves from inspiraling compact binaries:
       The quadrupole-moment term}
\author{Eric Poisson}
\address{Department of Physics, University of Guelph, Guelph,
         Ontario, N1G 2W1, Canada}
\date{Submitted to Physical Review D, September 12, 1997}
\maketitle

\begin{abstract}
\widetext
A rotating star's oblateness creates a deformation in the
gravitational field outside the star, which is measured by the
quadrupole-moment tensor. We consider the effect of the quadrupole 
moment on the orbital motion and rate of inspiral of a compact binary 
system, composed of neutron stars and/or black holes. We find that in 
the case of circular orbits, the quadrupole-monopole interaction 
affects the relation between orbital radius and angular velocity, 
and also the rate of inspiral, by a quantity of order $(v/c)^4$, 
where $v$ is the orbital velocity and $c$ the speed of light. 
\end{abstract}
\vskip 2pc]

\narrowtext

\section{Introduction and summary}

Inspiraling compact binaries, composed of neutron stars and/or 
black holes, are the most promising source of gravitational waves
for kilometer-scale interferometric detectors such as the American 
LIGO \cite{1} and the French-Italian VIRGO \cite{2}. The current 
construction of these detectors has motivated a lot of recent work 
on strategies to detect and measure such 
waves \cite{3,4,5,6,7,8,9,10,11}. In the course of this work, 
the need for extremely accurate predictions of the expected signal 
has repeatedly been demonstrated, as reliable model signals will 
be required for analyzing the data with the well-known technique of 
matched filtering \cite{12}. 

A large effort is currently underway to calculate inspiraling binary
waveforms to high order in a post-Newtonian approximation to the
exact laws of general relativity \cite{13,14,15}. This method is 
based on the assumption that the orbital motion is sufficiently 
slow that the waves can be expressed as a power series in $v$, 
the orbital velocity. Thus far, the waveforms have been calculated 
to order $v^5$ beyond the leading-order expressions, under the 
assumption that the spinning motion of the binary companions can 
be ignored \cite{14}. Rotational corrections to the waveforms have 
been computed only up to order $v^4$ \cite{16}.

Of particular importance for matched filtering is the phasing of the 
waves \cite{17}, which is determined by the rate of increase of the 
gravitational-wave frequency $f$ (equal to twice the orbital 
frequency) in response to the system's loss of energy and angular 
momentum to gravitational radiation. Assuming, as is usual, that the 
orbital motion is circular, this is given by \cite{5}
\begin{eqnarray}
\frac{df}{dt} &=& \frac{96}{5\pi {\cal M}^2}\, (\pi {\cal M} f)^{11/3}
\Biggl[ 1 - \biggl(\frac{743}{336} + \frac{11}{4}\, \eta \biggr)\, v^2
\nonumber \\ & & \mbox{}
+ (4\pi - \beta)\, v^3 
\nonumber \\ & & \mbox{}
+ \biggl( \frac{34103}{18144} + \frac{13661}{2016}\, \eta + 
\frac{59}{18}\, \eta^2 + \sigma \biggr)\, v^4
\nonumber \\ & & \mbox{}
+ O(v^5) \Biggr],
\label{1}
\end{eqnarray}
where we use units such that $c=G=1$. We have introduced a number of 
symbols. Let $m_1$ and $m_2$ denote the masses of the two companions, 
and let $M=m_1+m_2$ be the total mass and $\mu = m_1 m_2/M$ the reduced
mass. Then we define the chirp mass $\cal M$ and the mass ratio $\eta$
as
\begin{equation}
{\cal M} = \eta^{3/5} M, \qquad \eta = \mu/M.
\label{2}
\end{equation}
The orbital velocity $v$ is defined in terms of the gravitational-wave
frequency $f$ by
\begin{equation}
v = (\pi M f)^{1/3}.
\label{3}
\end{equation}
Finally, the quantities $\beta$ and $\sigma$ have to do with the spin
angular momenta of the companions, denoted $\bbox{s}_1$ and $\bbox{s}_2$.
The ``spin-orbit'' parameter is given by \cite{16}
\begin{equation}
\beta = \frac{1}{12} \sum_A \bigl[ 113(m_A/M)^2 + 75 \eta \bigr]\,
\bbox{\hat{L}} \cdot \bbox{\chi}_A,
\label{4}
\end{equation}
where $\bbox{\hat{L}}$ is the direction of the orbital angular
momentum, and $\bbox{\chi}_A = \bbox{s}_A/{m_A}^2$ is the 
dimensionless spin of companion $A$. (The index $A$ runs over the 
values 1 and 2. Here and throughout the paper, vectors and tensors 
are defined in three-dimensional flat space, and a hat indicates 
that the vector has a unit norm.) On the other hand, the ``spin-spin'' 
contribution to $\sigma$ is \cite{16}
\begin{equation}
\sigma_{\rm ss} = \frac{\eta}{48}\, \bigl[ 
-247 (\bbox{\chi}_1 \cdot \bbox{\chi}_2) 
+721 (\bbox{\hat{L}} \cdot \bbox{\chi}_1)
(\bbox{\hat{L}} \cdot \bbox{\chi}_2) \bigr].
\label{5}
\end{equation}

Our purpose in this paper is to derive an additional contribution 
to $\sigma$, namely $\sigma_{\rm qm}$, which is due to the quadrupole 
moments of the companions. The physical picture is the following. 
The spinning motion of companion $A$ creates a distortion in its 
mass distribution which, in turn, creates a distortion in the
gravitational field outside the star, measured by $Q_A^{ab}$, 
the (trecefree) quadrupole-moment tensor. The quadrupole term 
in the gravitational potential affects the orbital motion of the 
companions, and it affects also the emission of gravitational waves. 
This is the effect that we consider in this paper. It is important 
to understand that we are {\it not} considering the gravitational 
waves emitted by the time variations of the quadrupole moments 
$Q_A^{ab}$. Although such an effect exists (because of spin 
precession), it is much weaker (by an estimated factor of order 
$v^{16}$!) than the one considered here.

Assuming that the spinning body $A$ is axially symmetric about 
the direction of $\bbox{\hat{s}}_A$, the quadrupole-moment tensor 
can be expressed as
\begin{equation}
Q_A^{ab} = Q_A \bigl( \hat{s}^a \hat{s}^b - {\textstyle \frac{1}{3}}
\delta^{ab} \bigr),
\label{6}
\end{equation}
where $Q_A$ is the quadrupole-moment scalar. In Newtonian 
theory \cite{18}, this is given in terms of the mass density 
$\rho$ by $Q_A = \int_A \rho(\bbox{x})\, |\bbox{x}|^2\, 
P_2(\bbox{\hat{s}}\cdot \bbox{\hat{x}})\, d^3x$, where 
$P_2(x) = \frac{1}{2}(3x^2-1)$. In general relativity, $Q_A$ is 
defined in a coordinate-invariant manner in terms of the falloff 
behavior of the metric outside the star (see Ref.~\cite{19} and
references therein). The general relativistic definition reduces 
to the Newtonian one when the gravitational field is weak everywhere 
inside the star. 

In the following sections of this paper we calculate that the 
quadrupole-moment contribution to $\sigma$ is 
\begin{equation}
\sigma_{\rm qm} = - \frac{5}{2}\, \sum_A \frac{Q_A}{m_A M^2}\,
\Bigl[ 3 (\bbox{\bbox{\hat{L}} \cdot \hat{s}}_A)^2 - 1 \Bigr],
\label{7}
\end{equation}
where $Q_A$ is the general-relativistic quadrupole-moment scalar. 
This is the main result of this paper. To the best of the author's
knowledge, this contribution was never presented in the literature 
before, except in the specific context of a particle moving around a 
massive Kerr black hole \cite{20}. We note, however, that similar 
calculations were carried out independently, but not published, 
by Kidder \cite{21}. 

The quadrupole-monopole interaction responsible for $\sigma_{\rm qm}$
is a Newtonian effect which formally takes the appearance of a 
second post-Newtonian correction in $df/dt$. Indeed, apart from the 
facts that $Q_A$ is defined in a general-relativistic manner and
the Einstein quadrupole formula is used to compute the radiation, 
the calculation leading to Eq.~(\ref{7}) involves only Newtonian 
theory. In other words, although the gravitational field is not 
assumed to be weak inside the compact objects, the relative separation 
of the companions is assumed to be sufficiently large that the mutual 
gravitational potential will be small. Under such conditions, Newtonian 
theory can be used to describe the orbital motion, provided that the 
quadrupole moments are defined in a suitable, strong-field manner. 

It is well known that the quadrupole moment of a rotating black hole
is given by $Q_A = - {\chi_A}^2\, {m_A}^3$, where $\chi_A \equiv 
|\bbox{\chi}_A|$ is the dimensionless spin \cite{22}. (The minus sign 
reflects the oblateness of the black hole.) Quadrupole moments of 
realistic rotating neutron stars have been computed in Ref.~\cite{19}, 
where it is shown that for a neutron star of $1.4\, M_\odot$,
\begin{equation}
Q_A \simeq - a\, {\chi_A}^2\, {m_A}^3.
\label{8}
\end{equation}
Here, the parameter $a$ ranges from approximately 4 to 8 depending 
on the equation of state for neutron-star matter. Stiffer equations 
of state give larger values of $a$, and $a=1$ for a rotating black
hole. Inserting Eq.~(\ref{8}) into Eq.~(\ref{7}) shows that whatever 
the nature of the compact object (black hole or neutron star), 
$\sigma_{\rm ss}$ and $\sigma_{\rm qm}$ are of the same order of 
magnitude. For example, for two $1.4\, M_\odot$ neutron stars with 
$a=5.0$, we find $\sigma_{\rm ss} \in (-1.04, 1.04)$, while 
$\sigma_{\rm qm} \in (-2.64, 5.28)$. We have used the fact that
for such neutron stars, $\chi < 0.65$ \cite{19}.

The rest of the paper is organized as follows. In Sec.~II we integrate 
the Newtonian equations of motion for the centers of mass of the two 
companions, focusing on the motion over a time scale comparable with
the orbital period. In Sec.~III we consider the precessional motions
of the orbital and spin angular momenta, which occur over a time scale 
much longer than the orbital period. Finally, in Sec.~IV we calculate
the contribution to the rate of change of the gravitational-wave
frequency due to the quadrupole-monopole interaction. 

\section{Orbital motion}

The Newtonian equations of motion for the centers of mass of two 
bodies with masses $m_1$ and $m_2$ and quadrupole moments $Q_1$ 
and $Q_2$ can be derived from an effective one-body Lagrangian,
${\cal L} = \frac{1}{2} \mu \bbox{\dot{x}}^2 - V(\bbox{x})$,
where $\mu$ is the reduced mass, $\bbox{x} = \bbox{x}_2 - 
\bbox{x}_1$ the relative separation of the centers, and 
$\bbox{\dot{x}}$ the relative velocity. The gravitational 
potential is \cite{23}
\begin{equation}
V(\bbox{x}) = -\frac{\mu M}{r} - \frac{3}{2}\, 
\Bigl( m_1 Q_2^{ab} + m_2 Q_1^{ab} \Bigr) \frac{n_a n_b}{r^3},
\label{9}
\end{equation}
where $M$ is the total mass, $r \equiv |\bbox{x}|$, $\bbox{n} 
\equiv \bbox{\hat{x}} = \bbox{x}/r$, and $Q_A^{ab}$ is constructed 
from $Q_A$ as in Eq.~(\ref{6}). 

For what follows it is useful to define the dimensionless quantities
\begin{equation}
p_A = \frac{Q_A}{ m_A M^2 }
\label{10}
\end{equation}
and to introduce the angles $\alpha_A$ and $\beta_A$ such that
\begin{equation}
\bbox{\hat{s}}_A = (\sin\alpha_A \cos\beta_A, \sin\alpha_A \sin\beta_A,
\cos\alpha_A).
\label{11}
\end{equation}
Although these angles vary because of spin precession (see Sec.~III), 
this variation occurs on a time scale much longer than the orbital 
period. We shall therefore take them to be constant for the purpose of 
calculating the motion over a time scale comparable with the
orbital period. We also introduce the spherical coordinates 
$\{r,\theta,\phi\}$, such that $\bbox{x} = r\, \bbox{n}$ and
\begin{equation}
\bbox{n} = (\sin\theta \cos\phi, \sin\theta \sin\phi, \cos\theta).
\label{12}
\end{equation}
Substituting these relations, together with Eq.~(\ref{6}), into 
Eq.~(\ref{9}) yields
\begin{equation}
V(\bbox{x}) = -\frac{\mu M}{r} - \frac{\mu}{2}\, 
\biggl( \frac{M}{r} \biggr)^3 \sum_A p_A (3 \cos^2\gamma_A -1),
\label{13}
\end{equation}
where $\cos\gamma_A \equiv \bbox{\hat{s}}_A \cdot \bbox{n} =
\sin\alpha_A \sin\theta \cos(\phi-\beta_A) + \cos\alpha_A \cos\theta$.

The second term in Eq.~(\ref{13}) is smaller than the first by
a factor of order $(M/r)^2 = v^4$, and it can be treated as a
perturbation when integrating the equations of motion. Taking
the background motion to be a circular orbit in the equatorial
plane, we write $r(t) = r_0 [1 + \epsilon R(t)]$, 
$\theta(t) = \pi/2 + \epsilon \Theta(t)$, and
$\phi(t) = \Omega_0 t + \epsilon \Phi(t)$, where $r_0$ and
$\Omega_0 = M^{1/2} {r_0}^{-3/2}$ are the background radius and
angular velocity, respectively, and $\epsilon \equiv (M/r_0)^2$.
Substituting these relations into the equations of motion derived 
from $\cal L$ and linearizing with respect to $\epsilon$ leads to 
differential equations for the unknowns $R$, $\Theta$, and $\Phi$. 
Integration of these equations is straightforward, and we find 
\begin{eqnarray}
r(t) &=& r_0 \biggl[ 1 + \frac{3 \epsilon }{4}\, 
\sum_A p_A (3 \cos^2\alpha_A - 1) 
\nonumber \\ & & \mbox{}
+ \frac{\epsilon }{4}\, 
\sum_A p_A \sin^2\alpha_A \cos 2(\Omega_0 t - \beta_A) \biggr],
\label{14} \\
\theta(t) &=& \frac{\pi}{2} - \frac{3\epsilon}{4}\, \Omega_0 t
\sum_A p_A \sin 2\alpha_A \sin(\Omega_0 t - \beta_A),
\label{15} \\
\dot{\phi}(t) &=& \Omega_0 \biggl[ 1 - \frac{3\epsilon}{2}\,
\sum_A p_A (3 \cos^2\alpha_A - 1) 
\nonumber \\ & & \mbox{}
+ \frac{\epsilon}{4}\,
\sum_A p_A \sin^2\alpha_A \cos 2(\Omega_0 t - \beta_A) \biggr].
\label{16}
\end{eqnarray}
These equations show that the orbits cannot be circular in the
strict sense, although $r$ and $\dot{\phi}$ are constant after
averaging over an orbital period. As we shall see in Sec.~III,
the linear growth of $\theta$ signals the occurrence of orbital 
precession. Because we are concerned only with the motion over a 
time scale comparable with the orbital period, and because the 
precessional motion occurs over a much longer time scale, we shall 
ignore this issue here.

From Eqs.~(\ref{14}) and (\ref{16}) we find that the averaged 
orbital radius is given by
\begin{equation}
\langle r \rangle = r_0 \biggl[ 1 + \frac{3 \epsilon }{4}\, 
\sum_A p_A (3 \cos^2\alpha_A - 1) \biggr],
\label{17}
\end{equation}
while the averaged angular velocity is
\begin{equation}
\langle \dot{\phi} \rangle = \Omega_0 \biggl[ 1 - 
\frac{3\epsilon}{2}\, \sum_A p_A (3 \cos^2\alpha_A - 1) \biggr].
\label{18}
\end{equation}
Eliminating $r_0$ from these equations, we arrive at (we discard
all terms of order $\epsilon^2$ and higher)
\begin{equation}
\langle \dot{\phi} \rangle = 
\biggl( \frac{M}{\langle r \rangle^3} \biggr)^{1/2} \Biggl[
1 - \frac{3}{8}\, \biggr( \frac{M}{\langle r \rangle} \biggr)^2
\sum_A p_A (3 \cos^2\alpha_A - 1) \Biggr].
\label{19}
\end{equation}
Inverting this gives
\begin{equation}
\frac{M}{\langle r \rangle} = v^2 \biggl[ 1 + \frac{1}{4}\, 
v^4 \sum_A p_A (3 \cos^2\alpha_A - 1) \biggr],
\label{20}
\end{equation}
where $v \equiv (M \langle \dot{\phi} \rangle)^{1/3}$. 

We shall need also an expression for the orbital energy, 
given by $E = \frac{1}{2} \mu \bbox{\dot{x}} + V(\bbox{x})$. 
Substituting the relations (\ref{14})--(\ref{20}) and discarding 
all terms of order $\epsilon^2$ and higher, we find
\begin{equation}
E = - \frac{1}{2}\, \mu v^2 \biggl[ 1 + \frac{1}{2}\, v^4
\sum_A p_A (3 \cos^2\alpha_A - 1) \biggr].
\label{21}
\end{equation}
Equations (\ref{19})--(\ref{21}) describe the (averaged) orbital
motion over time scales comparable to the orbital period.

\section{Precessions}

It is well known that the Newtonian quadrupole-monopole interaction 
produces a precession of the orbital angular momentum, as well as a 
precession of the spin axis of each of the binary companions \cite{23}. 
We examine this effect here, as well as the precessional motions caused 
by the general relativistic spin-orbit and spin-spin interactions.

Calculating the torque associated with the potential (\ref{13}) and 
averaging over an orbital period leads to this equation describing
the orbital precession:
\begin{equation}
\langle \bbox{\dot{L}} \rangle = \bbox{\Omega}_{\rm op} \times \bbox{L},
\label{22}
\end{equation}
where the angular velocity of orbital precession is given by
\begin{equation}
\bbox{\Omega}_{\rm op} = \frac{3}{2}\, \biggl( \frac{M}{r} \biggr)^2\,
\Omega\, \sum_A p_A \cos\alpha_A\, \bbox{\hat{s}}_A.
\label{23}
\end{equation}
Here, $r \equiv \langle r \rangle$ is the orbital radius, and
$\Omega \equiv \langle \dot{\phi} \rangle$ is the orbital angular
velocity. Equation (\ref{23}) shows that the time scale for orbital
precession is longer than the orbital period by a factor of order 
$(r/M)^2$. Equations (\ref{22}) and (\ref{23}) also imply that 
$\cos\alpha_A \equiv \bbox{\hat{L}} \cdot \bbox{\hat{s}}_A$ does 
not stay constant during the motion. We note that Eq.~(\ref{15}) 
may be recovered by integrating Eq.~(\ref{22}) with $\bbox{\hat{L}}$ 
set equal to $\bbox{\hat{z}}$ on the right-hand side. 

On the other hand, treating companion $A$ as a rigid body under
the influence of the potential (\ref{13}) leads to this equation
describing the spin precession:
\begin{equation}
\langle \bbox{\dot{s}}_A \rangle = \bbox{\Omega}_{{\rm sp},A} 
\times \bbox{s}_A,
\label{24}
\end{equation}
where the angular velocity of spin precession is given by
\begin{equation}
\bbox{\Omega}_{{\rm sp},A} = \frac{3}{2}\, \frac{\mu M}{{m_A}^2}\, 
\biggl( \frac{M}{r} \biggr)^{3/2}\, \Omega\,
\frac{p_A \cos\alpha_A}{\chi_A}\, \bbox{\hat{L}} .
\label{25}
\end{equation}
This shows that the time scale for spin precession is longer than 
the orbital period by a factor of order $(r/M)^{3/2}$. It should 
be noted that for simplicity, we have not accounted for the time 
variation of $\cos\alpha_A$ in this calculation. 

The quadrupole-monopole interaction is not alone in inducing a
precession of $\bbox{L}$ and $\bbox{s}_A$. The general relativistic 
spin-orbit and spin-spin interactions also contribute to these 
precessions, and this was studied in detail in Ref.~\cite{24}. 
The spin-orbit contributions to the precessional angular velocities 
are given by
\begin{equation}
\bbox{\Omega}_{\rm op} = \frac{4m_1 + 3m_2}{2m_1 r^3}\, \bbox{s}_1
+ \frac{4m_2 + 3m_1}{2m_2 r^3}\, \bbox{s}_2
\label{26}
\end{equation}
and 
\begin{equation}
\bbox{\Omega}_{{\rm sp}, 1} = \frac{4 m_1 + 3m_2}{2 m_1 r^3}\, \bbox{L},
\label{27}
\end{equation}
with a similar equation holding for $\bbox{\Omega}_{{\rm sp}, 2}$. 
On the other hand, the spin-spin contributions are
\begin{equation}
\bbox{\Omega}_{\rm op} = - \frac{3}{2r^3 L}\, \Bigl[
(\bbox{s}_2 \cdot \bbox{\hat{L}})\, \bbox{s}_1 + 
(\bbox{s}_1 \cdot \bbox{\hat{L}})\, \bbox{s}_2 \Bigr],
\label{28}
\end{equation}
where $L \equiv |\bbox{L}| = \mu r^2 \Omega$, and
\begin{equation}
\bbox{\Omega}_{{\rm sp}, 1} = \frac{1}{2r^3}\, \Bigl[
\bbox{s}_2 - 3 (\bbox{s}_1 \cdot \bbox{\hat{L}})\, 
\bbox{\hat{L}} \Bigr],
\label{29}
\end{equation}
with a similar equation holding for $\bbox{\Omega}_{{\rm sp}, 2}$.

From these equations it is easy to see that the spin-orbit 
contribution to the precession dominates over both the spin-spin 
and quadrupole-monopole contributions. In fact,
\begin{equation}
\frac{\mbox{spin-orbit}}{\mbox{spin-spin}} \sim
\frac{\mbox{spin-orbit}}{\mbox{quadrupole-monopole}} \sim 
\biggl( \frac{r}{M} \biggr)^{1/2}.
\label{30}
\end{equation}
The main lesson here is that while the quadrupole-monopole
precession is entirely a Newtonian effect, it is a small one 
compared with the general relativistic spin-orbit precession, 
and it is of the same order of magnitude as the spin-spin 
precession.

These precessional motions are important, because they produce
modulations in the amplitude and phase of the gravitational
waves emitted by the binary system. The modulations produced by
the spin-orbit and spin-spin interactions were studied in detail 
in Ref.~\cite{24}. While it may be worthwhile to repeat this 
analysis so as to also incorporate the modulations due to the 
quadrupole-monopole interaction, it is doubtful that any of the 
conclusions would be affected. 

\section{Gravitational waves}

We now use the Einstein quadrupole formula \cite{25},
\begin{equation}
\dot{E} = - \frac{1}{5} \langle \overdots{Q}^{ab}\,
\overdots{Q}^{ab} \rangle,
\label{31}
\end{equation}
to calculate the rate of loss of orbital energy to 
gravitational radiation, for a binary system moving on a circular 
orbit of (averaged) radius $r \equiv \langle r \rangle$ and 
(averaged) angular velocity $\Omega \equiv \langle \dot{\phi} \rangle$, 
where $\langle r \rangle$ and $\langle \dot{\phi} \rangle$ are related 
by Eqs.~(\ref{19}) and (\ref{20}). The quadrupole moment of the centers 
of mass is given by $Q^{ab} = \sum_A m_A (x_A^a x_A^a - 
\frac{1}{3} {r_A}^2 \delta^{ab})$, where, 
in terms of the relative separation $\bbox{x}$, $\bbox{x}_1 = - 
m_2 \bbox{x} / M$ and $\bbox{x}_2 = m_1 \bbox{x}/ M$. Substituting 
$\bbox{x} = r(\cos\Omega t, \sin\Omega t, 0)$ into $Q^{ab}$, and
the result into Eq.~(\ref{31}), we find
\begin{equation}
\dot{E} = - \frac{32}{5}\, \mu^2 r^4 \Omega^6,
\label{32}
\end{equation}
irrespective of the relation between $r$ and $\Omega$. Using
Eq.~(\ref{20}) yields
\begin{equation}
\dot{E} = - \frac{32}{5}\, \biggl( \frac{\mu}{M} \biggr)^2
v^{10} \biggl[ 1 - v^4 \sum_A p_A (3\cos^2\alpha_A - 1) \biggr],
\label{33}
\end{equation}
where $v = (M\Omega)^{1/3}$.

The gravitational-wave frequency $f$ is equal to twice the orbital
frequency, so
\begin{equation}
\pi M f = M \Omega = v^3.
\label{34}
\end{equation}
The loss of orbital energy is accompanied by an increase in orbital 
frequency, and therefore, an increase in $f$. The rate of change of 
the gravitational-wave frequency is calculated as 
$\dot{f} = \dot{E}/(dE/df) = (3v^2/\pi M) \dot{E} / (dE/dv)$, 
where $E(v)$ is given by Eq.~(\ref{21}). This gives
\begin{equation}
\dot{f} = \frac{96 \mu}{5\pi M^3}\, v^{11} \biggl[ 1 -
\frac{5}{2} v^4 \sum_A p_A (3 \cos^2\alpha_A - 1) \biggr].
\label{35}
\end{equation}
The coefficient of the $O(v^4)$ term within the square brackets 
is the quadrupole-moment contribution to the parameter $\sigma$
appearing in Eq.~(\ref{1}). Taking into account the definitions
(\ref{10}) and (\ref{11}), this is the same as what was given
in Eq.~(\ref{7}).

\acknowledgments

This work was supported by the Natural Sciences and Engineering
Research Council of Canada. It is a pleasure to thank Bernie
Nickel for useful discussions on orbital precession.

\end{document}